\documentstyle[12pt,psfig,cite]{article}
\begin{document}
\renewcommand{\thefootnote}{\fnsymbol{footnote}}
\begin{titlepage}
\begin{flushright}
KEK-CP-039 \\
\end{flushright}

\parskip 0.8cm
\vspace{-0.5cm}
\Large
\begin{center}
2D Quantum Gravity \\
-Three States of Surfaces-
\end{center}

\normalsize
\begin{center}
A.Fujitsu \footnote[2]{E-mail address: a-fujitu@u-aizu.ac.jp}
 ,  
N.Tsuda \footnote[3]{E-mail address: ntsuda@theory.kek.jp}
 and
T.Yukawa \footnote[4]{E-mail address: yukawa@theory.kek.jp}
\end{center}

\normalsize
\begin{center}
{\it
$^{\dag}$ 
Information Systems and Technology Center The University of Aizu \\
Tsuruga, Ikki-machi, Aizu-Wakamatsu City Fukushima, 965-80, Japan

$^{\ddag}$ 
National Laboratory for High Energy Physics (KEK) \\
Tsukuba, Ibaraki 305, Japan

$^{\S}$ 
Coordination Center for Research and Education, \\
The Graduate University for Advanced Studies, \\
Hayama-cho, Miura-gun, Kanagawa 240-01, Japan \\
and \\
National Laboratory for High Energy Physics (KEK) \\
Tsukuba, Ibaraki 305, Japan
}
\end{center}

\begin{abstract}
\vspace{-0.6cm}
Two-dimensional random surfaces are studied numerically by the 
dynamical triangulation method. 
In order to generate various kinds of random surfaces, two higher 
derivative terms are added to the action.
The phases of surfaces in the two-dimensional parameter space are classified
into three states: {\it flat}, {\it crumpled surface}, and {\it branched 
polymer}.
In addition, there exists a special point (pure gravity) corresponding to 
the universal fractal surface.
A new probe to detect branched polymers is proposed, which makes use of 
the minbu(minimum neck baby universe) analysis.
This method can clearly distinguish the branched polymer phase from another
according to the sizes and arrangements of baby universes.
The size distribution of baby universes changes drastically at the 
transition point between the branched polymer and other kind of surface.
The phases of surfaces coupled with multi-Ising spins are studied in a 
similar manner.
\end{abstract}
\end{titlepage}

\section{Introduction}
Studies on two-dimensional surfaces are currently drawing much attention 
in various fields of science, such as physics, chemistry and biology.
In physics, for example, the statistical property of triangulated surfaces 
is under intensive investigations in the context of two-dimensional 
quantum gravity.
For a surface without coupled matter (pure gravity), a Monte-Carlo 
simulation by a dynamical triangulation(DT) method shows the fractal 
property, as predicted by such analytical theories as the Liouville 
field theory \cite{KPZ,DK,D}, the matrix model\cite{KKMWIK} and string 
field theory \cite{IK}. 
When matter, such as scalar fields, Ising spins or Potts models, 
is put on a surface it is expected that the surface becomes unstable 
toward forming a one-dimensional structure, like a branched polymer
\cite{BP,Cplx_Stru}. 
We are interested in generating and classifying various kinds of 
surfaces.
It is easy to imagine that there exist at least three kinds of surfaces, (
{\it flat, crumpled surface and branched polymer}), and wonder whether 
any surface can be classified into one of these three types.
We  would also like to investigate the nature of the transition between 
them.
Furthermore, if the fluctuation of a surface can be visualized, it will
be very useful for  understanding the properties of 
surfaces and mutual transitions.

One of the important quantities which characterize a surface generated by 
the dynamical triangulation method is the discretized local scalar curvature;
\begin{equation}
R_i = \frac{2\pi}{a^{2}}\frac{(6-q_i)}{q_i},
\label{eq:R}
\end{equation}
where $q_i$ (a coordination number) is the number of triangles sharing 
the vertex $i$, and $a^{2}$ is the area of an elementary triangle.
As is well-known, the surface of pure gravity fluctuates strongly, containing 
many branches and spikes.
A large positive curvature is localized at each tip of the spikes, and 
negative curvatures are condensed around the roots of the branches.
Two additional terms are included in the action in order to control 
the local fluctuation of the surfaces. 
One is $\beta \int \sqrt{g} R^{2} d^{2}x$; the other is $\alpha \int 
\sqrt{g} g^{\mu\nu} \partial_{\mu} R \partial_{\nu} R d^{2}x$ in continuous 
space, where $\alpha$ and $\beta$ are coupling constants.
The former affects on the local curvature of surfaces;{\it i.e.} for a large 
positive  $\beta$ it suppresses large fluctuations of $R$\cite{KN}, and 
as $\beta$ becomes negative the surfaces become crumpled or branched polymers, 
depending on the sign of $\alpha$.

Since the smallest neck size of a branch is limited by the lattice constant, 
it is significant to consider the minimum necks.
Only for the minimum neck can we locate its position uniquely.
By measuring the area distributions and connectivities of each baby universe 
separated by the minimum necks we can distinguish branched polymers from 
other kinds of surfaces.

This paper is organized as follows:
In the next section we briefly describe how the DT method is performed
with two higher derivative terms in the action. 
In Section 3 we give a short review of the method of minimum neck baby 
universes (or 'minbu'\cite{Minbu1}).
In Section 4 we propose three phases to classify random surfaces, and 
discuss the properties of each phase.
In Section 5 the transition between a crumpled surface (or a fractal surface) 
and a branched polymer is discussed, together with numerical simulation 
of random surfaces coupled with multi-Ising spins.
In the last section we give a summary of our numerical results and a 
discussion.

\section{Dynamical Triangulation with Higher \protect{\newline}
Derivative Terms}
A study of two-dimensional quantum gravity offers not only a simple model 
of Einstein gravity, but also a general framework for providing the 
universal property of two-dimensional surfaces.
Here, we employ the dynamical triangulation method \cite{DT} (DT), which is
known to give the correct answer, expected from analytic theories.  

In DT, calculations of the partition function are performed by replacing the 
path integral over the metric to the sum over possible triangulations of the 
two-dimensional surfaces by equilateral triangles. 
From the triangulation condition and the topological-invariance
the total number of i-simplices ($N_{i}$) is related as
\begin{equation}
3N_{2} = 2N_{1},
\end{equation}
\begin{equation}
N_{2} - N_{1} + N_{0} = \chi,
\end{equation}
where $\chi$ is the Euler number.

In practice we proceed with DT for a surface with an arbitrary genus as 
follows:
\begin{enumerate}
\item 
Initialization. 

Prepare equilateral 3-simplices(tetrahedrons), and put them together by 
gluing triangles face-to-face, forming a closed surface with a target genus.
Then repeat the barycentric subdivision by choosing a triangle randomly 
until the area of the universe is reaches to the target size.
This manipulation is equivalent to the so-called $(1,3)$ move.

\item
Thermalization of configurations.

For changing the geometry so as to bring the surface into thermal 
equilibrium, pick up a pare of neighbouring triangles randomly and flip the 
link shared by them(that is the flip-flop or $(2,2)$ move).
This move conserves the area and topology of the surface, and is known to 
be ergodic, {\it i.e.} any two configurations 
of a fixed area and topology are connected by the finite sequence of 
the $(2,2)$ moves.
The manifold conditions are always checked by requiring that any set of 
2-simplexes having a 0-simplex in common should constitute a 
combinatorial 1-ball$(S^{1})$.
We also restrict the geometry so as to not allow singular triangulations, 
such as self-energies and tadpoles, in the dual graph.
\end{enumerate}

For the case of pure gravity the action is independent of the geometry, 
and all graphs generated by the flip-flop moves are accepted with equal 
weight as those of a canonical ensemble.
In this case no dimensional parameter enters, and the surface is expected to 
be $fractal$.

Here, we introduce two additional terms in the action in order to produce 
various types of surfaces.
There is a correspondence between a continuous theory(characterized by 
the metric tensor $g_{\mu\nu}$) and a discretized theory(characterized by 
the coordination number $q_{i}$) through a relation
\begin{equation}
\int \sqrt{g} d^{2}x \cdots \Longleftrightarrow \frac{a^{2}}{3} 
\sum_{i=1}^{N_{0}} q_{i} \cdots.
\label{eq:correspond}
\end{equation}

Two new terms are expressed in the language of discretized theory by 
making use of eqs.(\ref{eq:R}) and (\ref{eq:correspond}) as
\begin{equation}
\left\langle \int \sqrt{g} R^{2} d^{2}x \right\rangle \cong 
\frac{a^{2}}{3}\left\langle \sum_{i}q_{i}R_{i}^{2} \right\rangle 
= \frac{4\pi^{2}}{3 a^{2}}\left\langle \sum_{i}
\frac{(6-q_{i})^{2}}{q_{i}}  \right\rangle
\label{eq:RR}
\end{equation}
and 
\begin{eqnarray}
\left\langle \int \sqrt{g} g^{\mu\nu} \partial_{\mu} R \partial_{\nu} R 
d^{2}x \right\rangle
& \cong & \frac{1}{\sqrt{3}} \left\langle \sum_{<i,j>} (R_{i}-R_{j})^{2} 
\right\rangle \nonumber \\
& = & \frac{1}{\sqrt{3}} \left\langle 2 \sum_{i} q_{i} R_{i}^{2}
-\frac{8\pi^{2}}{a^{4}}\sum_{<i,j>} (6-q_{i}) (6-q_{j}) / q_{i}q_{j}
\right\rangle,
\label{eq:DRDR}
\end{eqnarray}
where $\langle \cdots \rangle$ means the ensemble average and 
$\langle i,j \rangle$ indicates a nearest-neighbour pair of vertices.

These two terms control the order of the vertices.
The former favors a flat surface $({\it i.e.} \;\; q_{i} = 6)$, and the 
latter creates a correlation of curvatures between neighbouring pairs of 
vertices producing such as crumpled surfaces or branched polymers for 
positive or negative parameter $\alpha$, respectively.

From eq.(\ref{eq:DRDR}), since the $\int \sqrt{g} g^{\mu\nu} 
\partial_{\mu} R \partial_{\nu} R$ term contains the $R^{2}$ term, we thus 
rearrange it into the $\sum_{<i,j>} (6-q_{i}) (6-q_{j}) / 
q_{i}q_{j}$ term in numerical simulations instead of the r.h.s. of 
eq.(\ref{eq:DRDR}).
Then, the Lattice action is 
\begin{equation}
S = \beta_{Lattice} \sum_{i}\frac{(6-q_{i})^{2}}{q_{i}} + 
\alpha_{Lattice} \sum_{<i,j>} \frac{(6-q_{i})(6-q_{j})}{q_{i}q_{j}},
\end{equation}
where the script $Lattice$ indicates a dimensionless coupling,
\begin{equation}
\left\{
\begin{array}{ll}
\displaystyle{\alpha = - \frac{\sqrt{3}a^{4}}{8\pi^{2}} 
\alpha_{Lattice}}, \\[0.5cm]
\displaystyle{\beta  = \frac{3a^{2}}{4\pi^{2}}(\alpha_{Lattice} + 
\frac{1}{36} \beta_{Lattice})}.
\end{array}
\right.
\label{eq:corresp}
\end{equation}

Since these two higher derivative terms (eqs.(\ref{eq:RR}) and 
(\ref{eq:DRDR})) have dimensions of $a^{-2}$ and $a^{-4}$, respectively, 
they are irrelevant in the continuum limit.
The typical short-distance scale introduced by the $R^{2}$ term is 
$\sim \sqrt{\beta}$, within which the surfaces are controlled to be smooth, 
and we need to set $\beta > A$ in order to maintain an effect on surface 
significant over area $A$.

When both terms are included, it may be possible to construct an effective 
theory coupled with a scalar field by introducing an auxiliary field ($\chi$) 
and defining a theory for the case $\beta > 0$ by the action 
\begin{equation}
\int d^{2} \xi \sqrt{g} (\frac{-\alpha}{\beta^{2}} g^{\mu\nu} 
\partial_{\mu} \chi \partial_{\nu} \chi + \frac{1}{\beta} 
\chi^{2} - i \chi R),
\label{eq:scalarP}
\end{equation}
and for $\beta < 0$,
\begin{equation}
\int d^{2} \xi \sqrt{g} (\frac{-\alpha}{\beta^{2}} g^{\mu\nu} 
\partial_{\mu} \chi \partial_{\nu} \chi + \frac{1}{\beta} \chi^{2} 
+ \chi R).
\label{eq:scalarN}
\end{equation}
We then integrate out the auxiliary field formally, giving
\begin{equation}
\int d^{2} \xi \sqrt{g} R \frac{1}{\frac{1}{\beta} +  
\frac{\alpha}{\beta^{2}}\Delta} R,
\label{eq:RdR}
\end{equation}
obtaining the effective action, 
\begin{equation}
S = \int \sqrt{g} d^{2}\xi (\alpha g^{\mu\nu} \partial_{\mu} R \partial_{\nu} 
R + \beta R^{2} ), \;\;\;\;\;\; \mbox{for} \; |\alpha| \ll |\beta|,
\end{equation}
in a perturbative expansion using the parameters $\frac{\alpha}{\beta}$.
In order to guarantee the perturbative expansion of eq.(\ref{eq:RdR}), 
$\alpha$ and $\beta$ must be restricted to $|\alpha| \ll |\beta|$
\footnote
{
Due to the restriction of parameters($\alpha,\beta$), 
our numerical results cannot include the so-called Feigin-Fuchs-type 
action, $\int d^{2} \xi \sqrt{g} (\alpha g^{\mu \nu} \partial_{\mu} \chi 
\partial_{\nu} \chi - i\chi R)$.
}.
It is important to notice that the ratio of the couplings, $\frac{\alpha}
{\beta^{2}}$, can be fixed to be finite, while both parameters $\alpha$ and 
$\beta$ are taken to have infinite simultaneity in order to keep the effects 
significant in the continuum limit.
Since the signs of the kinetic terms of eqs.(\ref{eq:scalarP}) and 
(\ref{eq:scalarN}) must be positive, it is required that $\alpha < 0$.
We again discuss the validity of the model of eq.(\ref{eq:scalarP}) in 
subsection $4.1$.

\section{Analysis of the baby universes}
In order to discuss the stability (or instability) of two-dimensional 
surfaces, we consider the so-called minimum-neck baby universes, as were 
mentioned in section 1.
A minimum-neck baby universe is defined as a connected sub part of a surface 
whose area must be less than half of the total area, and the neck is 
constructed by three links which are closed, non-self intersecting.
A mother universe is defined as a minimum-neck universe whose area is 
greater than half the total area.

Suppose two kinds of surface configurations(in Fig.1(a)): one is a 
non-branched sphere, whose volume is $A$; the other is branched sphere 
which has a baby universe(volume $B$).
Once the latter configuration is favorable, all universes branch out 
boundlessly.
Then, a bottle neck becomes a minimum-length neck, which leads to 
lattice-branched polymers.
Instabilities make the surfaces branched, and simultaneously the 
mother universe disappears.
Whether there exists a mother universe or not is one of the 
differences between branched polymers and other kinds of surfaces.

It is easy to show that the relationship between of the string 
susceptibility to the branched polymer instability\cite{Kawai}, 
for the partition function with the total area constrained to $A$ given by 
\begin{equation}
Z(A) \sim \kappa^{A} A^{\gamma - 3}.
\end{equation}
For the case of a branched configuration (Fig.1(a)), we can estimate a 
lower bound for this contribution, 
\begin{equation}
\int_{\frac{A}{3}}^{\frac{2A}{3}} dB Z(B)Z(A-B)B(A-B) \sim 
\kappa^{A} A^{2 \gamma - 3}.
\end{equation}
Since Z[A] is the total number of distinct graphs with $A$ triangles 
and $S^{2}$ topology, $BZ[B]$ is the total number of distinct graphs with $B$ 
triangles and $D^{2}$ topology for a boundary loop length of 3.

The dominance of the branched polymer configuration corresponds to the case of 
$\gamma > 0$. 
Thus, it is natural to recognize that a positive $\gamma$ corresponds to the 
case $c>1$.
So far, since there have been no exactly solvable models with $c>1$, it is 
important to measure the actual minbu distributions for various kinds of DT 
surfaces in order to understand the dynamics of a 2d surface.

In practice, it is easy to identify all of the minimum necks, and their 
connectivities on real dynamically triangulated graphs without 
self-energies and tadpoles in a dual graph.
The procedures are as follows:
\begin{itemize}
\item[(I)]
Pick up one link. There are then two vertices at both edges of 
this link; each vertex connects with many other links.
Among them(links) a common vertex may be contained; unless these 
three links form a $2$-simplex (${\it i.e.}$ a triangle) we call these 
closed links a minimum neck. 
A smaller part of the two universes separated by this minimum neck is a minbu.
Repeat this process until all of the links are checked.
This is applicable to the configurations of a nest of baby universes.

\item[(II)]
The connectivities of each baby universe are easily found. 
The key property is that a neck always separates universe into two parts for 
a spherical topology.
First, pick up an arbitrary neck; then, two separated universes ({$U1,U2$}) 
can be identified.
Second, choose a neck which belongs to universe $U1$(or $U2$); then, 
$U1$(or $U2$) can be decomposed into two sub parts: {$U11,U12$}(or 
{$U21,U22$}).
Repeat until all of the necks have been checked.
After this, it is straightforward to know the area distributions of all 
universes.
\end{itemize}

The minbu analysis can also visualize a triangulated surface as a network 
of baby universes connected by minimum necks (Fig.1(b)).
In the next section we apply this analysis in order to classify the surfaces.

A measurement of the minbu distributions provides the string susceptibility 
($\gamma_{s}$) with high accuracy\cite{Minbu2}. 
The probability of finding a minbu with area $B$ in a universe having a total 
area of $A$ is given by
\begin{equation}
P_{A}(B) = \frac{3 B Z[B] (A-B)Z[A-B]}{Z[A]}
\label{eq:Prob_Minbu}
\end{equation}

In order to obtain the frequency of finding a minbu ($N_{A}(B)$) with volume 
$B$ for the case of pure gravity, we substitute the KPZ formula 
into eq.(\ref{eq:Prob_Minbu}), giving 
\begin{equation}
N_{A}(B) \sim A^{\gamma_{s} - 1} \left(\frac{B}{A} \right)^{\gamma_{s} - 2}
\left\{ 1-\left(\frac{B}{A}\right)\right\}^{\gamma_{s} - 2},
\label{eq:minbu}
\end{equation}
where $\gamma_{s}=-\frac{1}{2}$. 

It is easy to extract the $\gamma_{s}$ from numerical data when the KPZ 
formula is applicable.
It should be remarked that in the case of $c \ne 0$ the finite size 
corrections for eq.(\ref{eq:minbu}) become important.
Even in the case of one Ising spin coupled to gravity (correspond to 
$c=\frac{1}{2}$), the finite size corrections are not negligible, and without 
a proper correction it is not possible to define the correct 
$\gamma_{s}(=\frac{1}{3})$ with the same accuracy as in the case of 
$c=0$ \cite{Harri_Yukawa}.

\begin{figure}
\vspace*{0cm}
\centerline{
\psfig{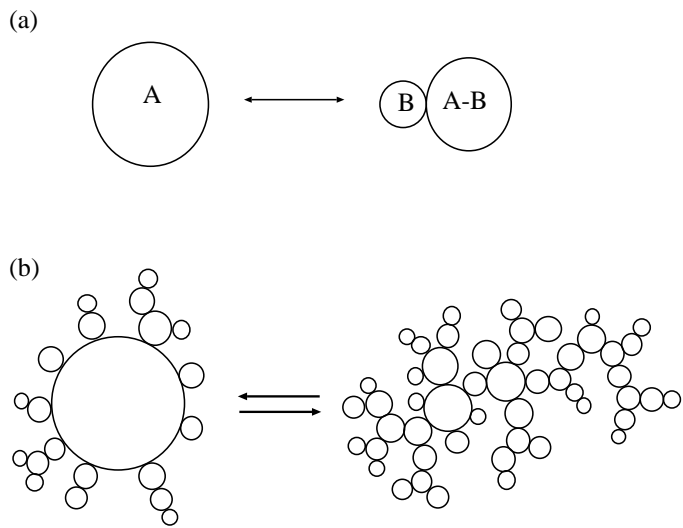}} 
\caption
{
$(a)$ Non branched surface with volume A and a branched surface
which includes one baby universe with volume B.
$(b)$ A surface generated by the DT method visualized as a network of 
baby universes.
One of the examples of the transition between fractal(or crumpled) surfaces 
to branched polymers is also visualized. 
}
\label{fig:minbu_neck}
\vspace{0.5cm}
\end{figure}

\section{Phase Diagram}
Let us classify typical states of surfaces 
generated by the two additional higher derivative terms in the action.
We would like to propose that any surfaces are constituted by a branched 
polymer, a crumpled surface and a flat one.
We can intuitively explain how these kinds of surfaces are created. 
The term $\sum_{i} (6-q_{i})^{2}$ with a positive coefficient favors 
$q_{i} = 6$, $i.e.$ a flat surface.
On the other hands, the term $\sum_{i} (6-q_{i})^{2}$ with a negative 
coefficient favors crumple surfaces rather than branched polymers. 
The term $\sum_{<i,j>} \frac{(6-q_{i})(6-q_{j})}{q_{i}q_{j}}$ with a positive 
coefficient also makes surfaces crumple, because the opposite-sign 
curvatures of adjacent vertices become attractive.
On the contrary the term $\sum_{<i,j>} \frac{(6-q_{i})(6-q_{j})}
{q_{i}q_{j}}$ with a negative coefficient makes surface branched polymers, 
because same-sign curvatures are attractive.
For the root(the loop of negative curvatures) and tip(a hemisphere of 
positive curvatures), those curvatures having the same sign need to condense 
to form a branched polymer. 

In order to classify the surface, we have observed $Q_{2}$ and $Q_{3}$ which 
are ensemble-averaged quantities ($cf$ eqs.(\ref{eq:RR}) and (\ref{eq:DRDR})), 
defined on the lattice as follows:
\begin{equation}
Q_{2} \equiv \langle \sum_{i}\frac{(6-q_{i})^{2}}{q_{i}} \rangle / 
N_{2}, 
\label{eq:Q2_def}
\end{equation}
\begin{equation}
Q_{3} \equiv \langle \sum_{<i,j>} 
\frac{(6-q_{i})(6-q_{j})}{q_{i}q_{j}} \rangle / N_{2}.
\label{eq:Q3_def}
\end{equation}

\begin{figure}
\vspace*{0cm}
\centerline{
\psfig{file=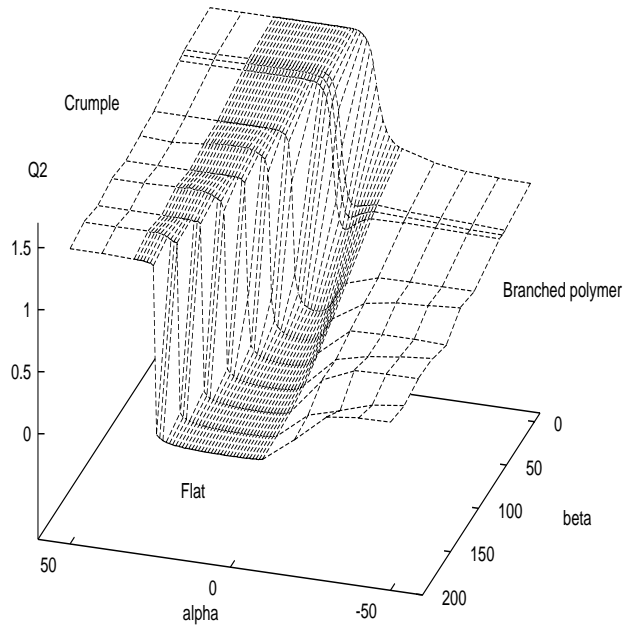,height=10cm,width=10cm,angle=-90}} 
\caption
{
Plot of $Q_{2}$ versus $\alpha$ and $\beta$.
}
\label{fig:Q2}
\vspace{0cm}
\end{figure}

\begin{figure}
\vspace*{0cm}
\centerline{
\psfig{file=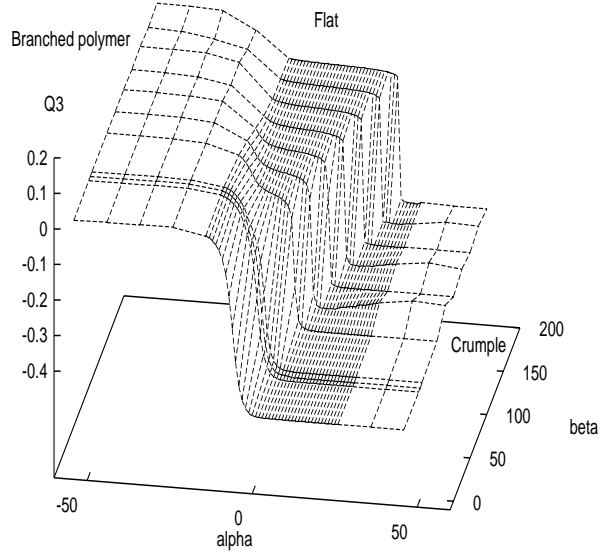,height=10cm,width=10cm,angle=-90}} 
\caption
{
Plot of $Q_{3}$ versus $\alpha$ and $\beta$.
}
\label{fig:Q3}
\vspace{0cm}
\end{figure}

Figs.\ref{fig:Q2} and \ref{fig:Q3} clearly show that there are three 
plateaus, which indicate that the DT surfaces can be classified into three 
characteristic states ($i.e.$ crumpled, flat and branched polymer) depending 
on the two parameters which we have employed.

We have also measured the size-distribution and arrangements of all 
minimum-neck baby universes on these three plateaus.
The features of these three surfaces are itemized as follows:
\begin{enumerate}
\item
 {\it Crumpled Surface} ($\alpha > 0, \beta < \alpha$) : 

This surface looks jagged.
The value of $Q_{2}$ becomes much larger than in the flat case, and that of 
$Q_{3}$ becomes negative.
A mother universe exists whose volume reaches the order of $60 \%$ 
or $70 \%$ of the total area, although the growth of large branches is 
highly suppressed(see Fig.\ref{fig:NumberofBranch}).
All baby universes have a small volume ($\sim {\cal O}(1)$).
The number of minimum necks is about $8 \sim 9 \%$ of the total 
number of triangles.
\item
 {\it Flat Surface} ($\beta > |\alpha|$) :

Almost all fluctuations are suppressed ($i.e.$ $Q_{2} \approx Q_{3} 
\approx 0$), and the fractal dimension becomes about $2.0$.
\item
 {\it Branched Polymer} ($\alpha < 0 , \beta < |\alpha|$) :

One of the typical branched polymers is illustrated by the r.h.s. of 
Fig.\ref{fig:minbu_neck}$(b)$.
The mother universe completely disappears.
All baby universes are very small ($\sim {\cal O}(1)$).
The number of minimum necks is about $30 \sim 40 \%$ of the total number 
of triangles.
\end{enumerate}

\begin{figure}
\vspace*{0cm}
\centerline{
\psfig{file=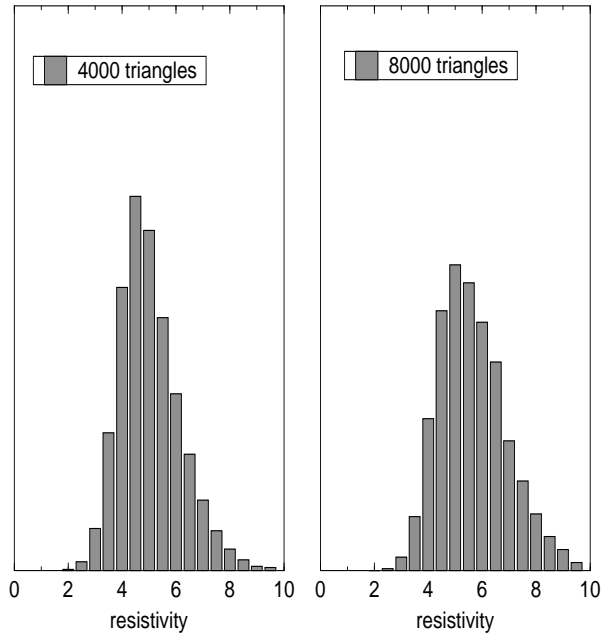,height=9cm,width=9cm,angle=0}} 
\caption
{
Normalized histogram of the resistivities\protect{\cite{Cplx_Stru}} for the 
case of branched-like surfaces: $\alpha = -10,\beta = 0$.
}
\label{fig:Comp_Resist}
\vspace{0cm}
\end{figure}

\begin{figure}
\vspace*{0cm}
\centerline{
\psfig{file=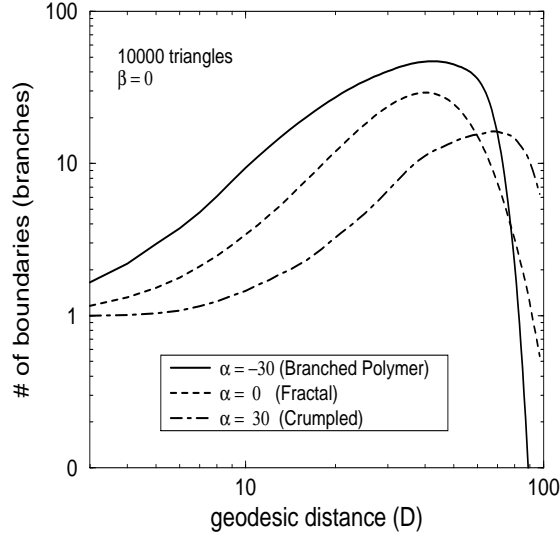,height=8cm,width=8cm,angle=0}} 
\vspace{-0.3cm}
\caption
{
Distributions of the number of boundaries (branches) versus the geodesic 
distance for various states with double-log scales.
The solid line represents branched polymer states, the dotted line represents 
the case of pure gravity (fractal states) and the dashed-dotted line represents 
crumpled states.
}
\label{fig:NumberofBranch}
\vspace{0cm}
\end{figure}

At the $N_{2} \to \infty$ limit of the branched polymer its surface can 
no longer be regarded as being a smooth continuum medium.
In order to check this, we measured the resistivities (correspond to the 
complex structures defined on DT surfaces) of branched 
polymers based on ref.\cite{Cplx_Stru}.
If there exists a smooth continuum limit of this surface, the peaks of 
resistivity become narrower as the size increases.
Fig.\ref{fig:Comp_Resist} is a plot of a normalized histogram of the 
resistivities for branched polymers.
As has been expected above, Fig.\ref{fig:Comp_Resist} shows that 
there is no sharpening of the peaks as the size increases.
That indicates that a branched polymer has no continuum limit.

Furthermore we point out the characteristic property based on an 
intrinsic geometrical point of view of these three states, that is, the 
distributions of the number of boundaries (branches) versus the geodesic 
distances ($D$). 
A definition of the boundaries (branches) is given in the next subsection 
($4.2$).
Fig.\ref{fig:NumberofBranch} supports the intuitive pictures that we 
mentioned above.
Especially, the dashed-dotted line (correspond to the crumpled space-time) 
indicates that the large branch structures are highly suppressed.
This tendency coincides with that of a case of three-dimensional simplicial 
gravity\cite{3DSG}.

\subsection{Flat states ($R^{2}$-gravity)}
\begin{figure}
\vspace*{-0.5cm}
\centerline{
\psfig{file=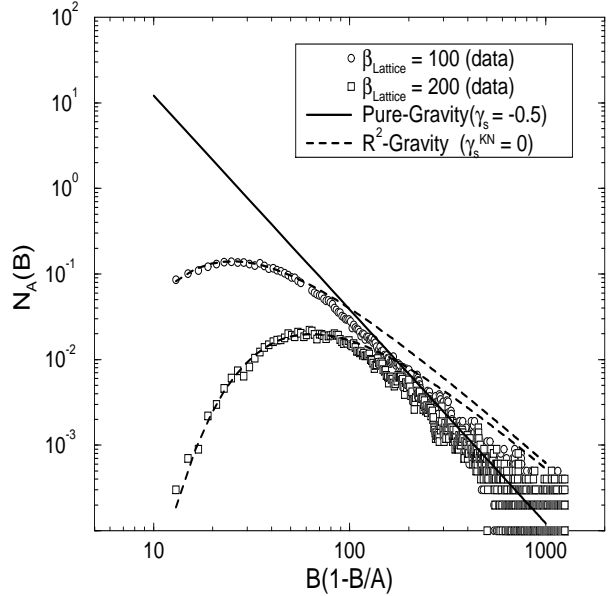,height=9cm,width=9cm}} 
\vspace{-0.5cm}
\caption
{
Minbu area distributions ($N_{A}(B)$) for $R^{2}-$gravity with 
topology $S^{2}$ with the double-log scales. 
The total number of triangles is 5,000.
$B$ represents the area of a baby universe.
The solid line represents the minbu area distributions for the pure-gravity 
case (eq.(\protect{\ref{eq:minbu}})).
The two dashed lines show the fit curve predicted by $R^{2}-$gravity \cite{KN} 
with $\beta_{Lattice}$ = 100 and 200.
}
\label{fig:Minbu_Comp_5K}
\vspace{0cm}
\end{figure}

Here, we consider in detail the almost flat states of DT surfaces.
At first, we investigate the minbu distributions for a large positive $\beta$
($\alpha$ = 0) \cite{JT}.
In ref.\cite{KN} the partition function and the string susceptibility have 
been estimated to be the following asymptotic form for $\frac{A}{\beta} \to 0$ 
and $A \to \infty$:
\begin{equation}
Z^{KN} \sim A^{\gamma^{KN}_{s}-3} \exp(-64\pi^{2}(1-h)^{2}\cdot \frac{\beta}
{A})e^{-\mu A},
\label{eq:KN_partition}
\end{equation}
where  
\begin{equation}
\gamma^{KN}_{s}(\chi, \frac{A}{\beta}) = 2 + \frac{d-12}{12}\chi 
+ \frac{d}{768 \pi^{2}}\frac{A}{\beta}.
\label{eq:Gamma_KN1}
\end{equation}
Here, $\chi$ is the Euler number of this surface, and $d$ is the central 
charge of the matter fields.
For the case of $d=0$ and $\chi=2 (\mbox{sphere}:S^{2})$, $\gamma^{KN}_{s}$ 
becomes zero.
In Fig.\ref{fig:Minbu_Comp_5K} we plot and compare the minbu distributions 
($N_{A}(B)$) versus $B(1-\frac{B}{A})$ with the double-log scale for 
$\beta_{Lattice} = 100$ and $200$ with a total number of triangles 5000.
Since the $R^{2}$ term affects only on the short-distance property of surfaces, 
it is expected that eq.(\ref{eq:Gamma_KN1}) is adaptable for a relatively small 
minbu region.
Fig.\ref{fig:Minbu_Comp_5K} shows a good agreement with numerical data(open 
circle and square), and an asymptotic formula (dashed-line) which is obtained 
by substituting eq.(\ref{eq:KN_partition}) into eq.(\ref{eq:Prob_Minbu}) 
within the small $B$ regions
\footnote
{
Since this asymptotic form is only assumed to be valid for $B \gg 1$, we 
should consider the simplest type of correction for $B$ $i.e.$ $B^{\gamma -2}
\to B^{\gamma -2}(1 + \frac{\tilde{C}}{B} + \cdots)$ \cite{Minbu2}.
Two dashed-lines in Fig.\ref{fig:Minbu_Comp_5K} are as follows: 
\begin{equation}
\ln(N_{A}(B)) = C_{0} -2 \ln\{B(1-\frac{B}{A})\} + 
\frac{C_{1}}{B(1-\frac{B}{A})},
\end{equation}
where $C_{0}$ and $C_{1}$ are fit parameters.
}.
We note that the data points approach the line predicted for the pure-gravity 
case ($\gamma_{s} = -\frac{1}{2}$) in large minbu regions. 
This indicates that the $R^{2}$ term is irrelevant with respect to long-range 
structures.

When we consider a part of a flat parameter region, 
$0 < \alpha_{Lattice} \ll \beta_{Lattice}$, the model of 
eq.(\ref{eq:scalarP}) corresponds to our numerical simulations.
The string susceptibility ($\gamma^{KN}_{s}$) of the model (\ref{eq:scalarP}) 
was also obtained by ref.\cite{KN} in the asymptotic form for 
$\frac{A}{\beta}\to 0$ and $A \to \infty$,
\begin{equation}
\gamma^{KN}_{s} = 2 - \left(\frac{24 + \eta}{12} + \frac{1}{32\pi} \cdot 
\frac{\beta^{2}}{\alpha} \sqrt{\frac{24 + \eta}{\eta}} \right) 
\frac{\chi}{\alpha_{-}} - \frac{d}{8 \pi \alpha_{-} \eta} \cdot 
\frac{\beta^{2}}{\alpha} \sqrt{\frac{24 + \eta}{\eta}} \cdot \frac{A}{\beta},
\label{eq:Gamma_KN2}
\end{equation}
where $\eta \equiv - \frac{3}{8\pi} \cdot \frac{\beta^{2}}{\alpha} - d$ and 
$\alpha_{-} \equiv \frac{1}{12} (24 + \eta - \sqrt{\eta (24 + \eta)})$.
It is remarkable that the $\gamma^{KN}_{s}$ of eq.(\ref{eq:Gamma_KN2}) does 
not depend upon $\alpha$, and is the same as $\gamma_{s}$ of 
eq.(\ref{eq:Gamma_KN1}) when $d=0$.
The plateau(corresponds to $|\alpha| < \beta$) observed in Figs.\ref{fig:Q2} 
and \ref{fig:Q3} indicates the $\alpha$-independence 
of $\gamma^{KN}_{s}$.

As reported in ref.\cite{Tsuda_Yukawa,ITY}, the transition from the fractal 
configuration to the flat configuration was a cross-over. 
It can be realized by varying $\beta$ with $\alpha = 0$ in our case.
In Fig.\ref{fig:R2} we plot $Log(Q_{2})$ versus $Log(\beta_{Lattice})$, 
comparing the semiclassical calculation (see appendix A) and numerical 
results with the total number of triangles ($20,000$).
A semi-classical analysis shows the excellent agreement with the numerical 
data at $\beta \approx 0$ and large $\beta$.

\begin{figure}
\vspace*{-0.5cm}
\centerline{
\psfig{file=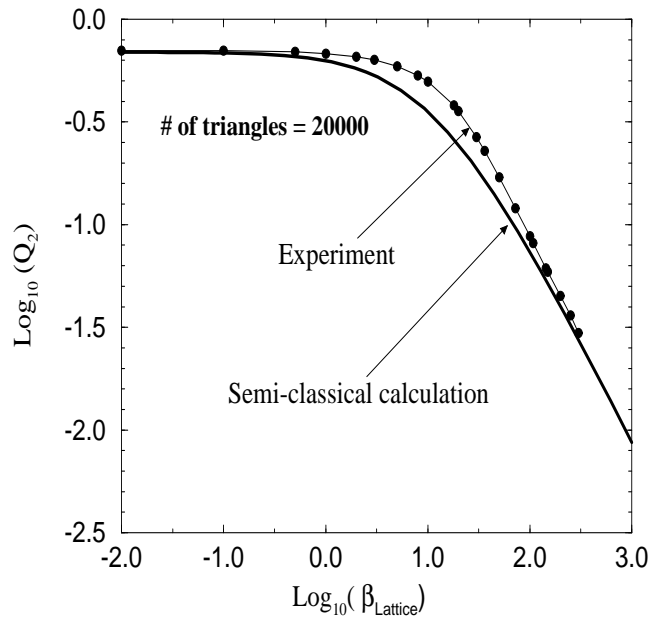,height=9cm,width=9cm}} 
\caption
{
Plot of $Log_{10}(Q_{2})$ versus $Log_{10}(\beta_{Lattice})$, comparing 
the semi-classical calculation and numerical results with a total number of 
triangles of $20,000$.
The filled circles represent the numerical data with $\alpha_{Lattice} = 0$ 
and $\beta_{Lattice} = 0.01 \sim 300$; the thin line is a guide.
The thick line represents a semi-classical analysis. (See Appendix A).
}
\label{fig:R2}
\vspace{0cm}
\end{figure}

\subsection{Fractal structures at $\alpha \sim \beta \sim 0$}
\begin{figure}
\vspace*{0cm}
\centerline{
\psfig{file=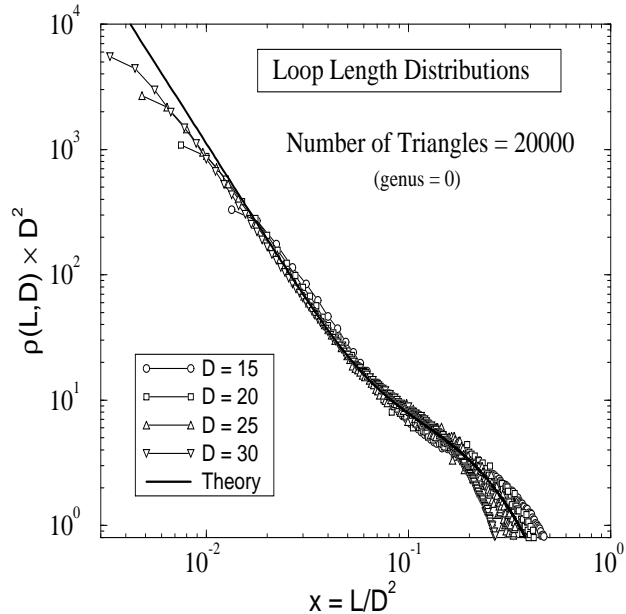,height=9cm,width=9cm}} 
\vspace{-0.5cm}
\caption
{
Loop-length distributions with the double-log scales. 
The total number of triangles is 20,000, and is averaged over 1000 
configurations. $x = {L}/{D^2}$ is a scaling variable, where $D$ 
represents the geodesic distance (it was measured at steps $D=15,20,25,30$) 
and $L$ represents each loop length at step $D$. 
The small circles, quadrangles and triangles indicate the results of 
numerical simulations; the solid line indicates the theoretical curve 
predicted by string field theory\cite{KKMWIK}.
}
\label{fig:LLD20K}
\vspace{-0.2cm}
\end{figure}

The most significant transition point is at ($\alpha \sim \beta \sim 0$).
Here, no scale parameters exist, and the surfaces are expected to become a 
fractal.
We now define the intrinsic geometry using the concept of a geodesic distance 
between two triangles on a DT surface.
Suppose that a disk which is covered within $D$ steps from some reference 
triangle. 
Because of the branching of the surface, a disc is not always 
simply-connected, and there usually appear many boundaries which consist 
of $S^{1}$ loops in this disc.
In order to show the fractal properties of the DT surface, the loop-length 
distribution function ($\rho(L,D)$) is measured by counting the number of 
boundary loops with the length $L$ which make boundaries of the area covered 
within $D$ steps.
In ref.\cite{KKMWIK} it has been predicted that the loop-length distribution 
is a function of the scaling variable, $x=l/d^{2}$, in the continuum limit for 
pure gravity, where $l$ is the loop length and $d$ is a distance defined on 
the continuum surfaces. 
For simplicity we do not distinguish continuous variables ($l,d$) from 
lattice variables ($L,D$).
In Fig.\ref{fig:LLD20K} the results of our simulation for various distances 
are compared to the theory for a surface with the size of $20,000$ triangles.
The distributions with different distances show an excellent agreement with 
each other and the shapes of the numerical data and the theoretical curve 
are quite similar.
At the same time, artifacts of a lattice at small $x$ and the finite-size 
effects at large $x$ are observed.
We have also investigated for several sizes of surfaces, $i.e.$ $100,000$
and $400,000$ triangles \cite{Tsuda_Yukawa}, and the same quantities have 
been given(but low statistics).
These excellent agreements with the numerical results and the analytical 
approach make it clear that the DT surface becomes fractal in the sense 
that sections of the surface at different distances from a given point look 
exactly the same after a proper rescaling of the loop lengths.
The loop-length distribution function comprises two types of loops 
corresponding to the exponent: baby loops and a mother loop.
The baby loops dominate the loop length, and give the fractal dimension $4$; 
however, their averaged length is on the order of the cut-off length, and 
they are non-universal, and it cannot be considered as the proper fractal 
dimension of this surface.
There always exists one mother loop with length proportional to $D^{2}$, which 
has a fractal dimension of 3.
In our earlier simulations\cite{MyThesis}, we found that the naive fractal 
dimension reached about $3.5$ with $400,000$ triangles.

\section{Transition to a Branched Polymer}
The $\alpha$ term ($\alpha > 0$) favors the attraction of vertices with the 
opposite-sign curvature.
If this coefficient still becomes larger, this effect exceeds the $R^{2}$ term 
effect, and the surface crumples.
On the other hand, a negative coefficient ($\alpha < 0$) makes a transition 
to the branched polymer.
In Fig.\ref{fig:minbu_neck}$(b)$ we illustrate one such kind of transition.

\begin{figure}
\vspace*{0cm}
\centerline{
\psfig{file=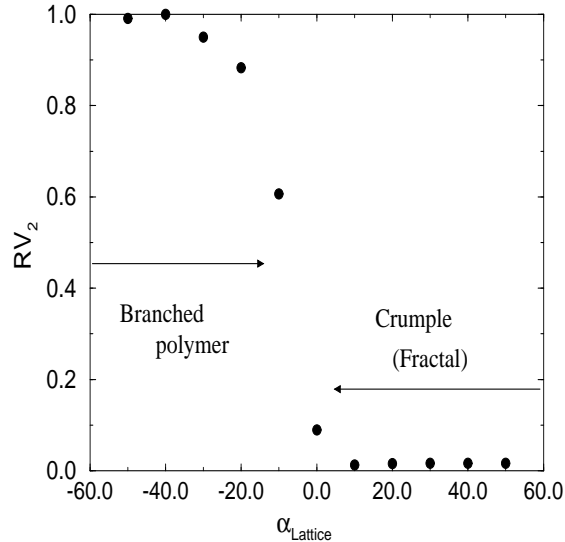,height=8cm,width=8cm}} 
\caption
{
$RV_{2}$ plotted versus $\alpha_{Lattice}$.
The total number of triangles is 2,000, and $\beta=0$.
The statistical errors were estimated to be about $10$\%.
}
\label{fig:Ratio_A1_A2}
\vspace{0cm}
\end{figure}

The important point to note is the volume distributions of a mother universe 
and a baby one.
For fractal surfaces(corresponding to the pure-gravity case), a mother 
universe has a large volume $i.e.$ $\sim$ half of the total volume(
${\cal O}(N_{2})$), and the sizes of the other baby universes are very small
(${\cal O}(1)$). 
On the other hand, for branched polymer configurations all of the baby 
universes have small sizes and no mother universe exists. 
We therefore propose a new order parameter($RV_{1,2}$) for the transition to 
the branched polymer,
\begin{equation}
RV_{1} \equiv \frac{V_{max}}{V_{total}} \;\; \mbox{or} \;\; 
RV_{2} \equiv \frac{V_{second}}{V_{max}},
\end{equation}
where $V_{max}$ represents the maximum area among all of the universes in 
this surface, and $V_{second}$ represents a secondary size universe.
Indeed, we examined that $RV_{1,2}$ works well as an indicator of such 
kinds of transitions in numerical simulations.
Fig.\ref{fig:Ratio_A1_A2} shows the $RV_{2}$ versus $\alpha_{Lattice}$ with 
the range: $[-50,50]$.
A clear transition can be seen from the branched polymer to crumpled or 
fractal surfaces.

\begin{figure}
\vspace*{-1.0cm}
\centerline{
\psfig{file=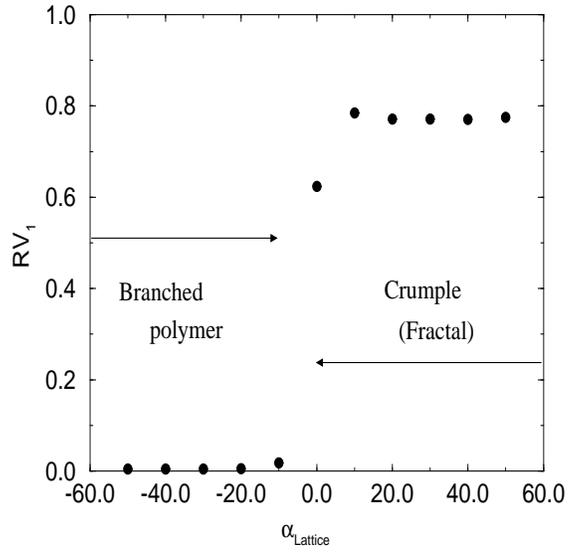,height=8cm,width=8cm}} 
\caption
{
$RV_{1}$ plotted versus $\alpha_{Lattice}$.
The other conditions are the same as in Fig.9.
}
\label{fig:MinbuMaxVol}
\vspace{-0.5cm}
\end{figure}

\begin{figure}
\vspace{-1cm}
\centerline{
\psfig{file=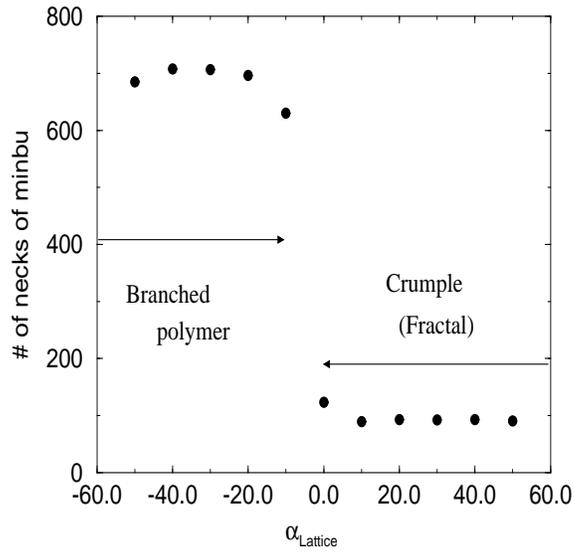,height=8cm,width=8cm}} 
\caption
{
Total number of minimum necks plotted versus $\alpha_{Lattice}$.
The numerical data are indicated by filled circles.
The other conditions are the same as in Fig.9.
}
\label{fig:MinbuNeckTotNumber}
\vspace{-0.5cm}
\end{figure}

In Figs.\ref{fig:MinbuMaxVol} and \ref{fig:MinbuNeckTotNumber} we also show 
$RV_{1}$, as well as the total number of minimum necks versus 
$\alpha_{Lattice}$, respectively. 
These data (Fig.\ref{fig:Ratio_A1_A2},\ref{fig:MinbuMaxVol} 
and \ref{fig:MinbuNeckTotNumber}) suggest that the DT surfaces become a 
branched polymer for $\alpha < -10$.
As has been mentioned, our numerical systems correspond to the effective 
theory coupled with a scalar field; also $\alpha$ must be negative 
($i.e.$ $\alpha_{Lattice}$ must be positive). 
Then, the branched polymer stands for the appearance of some instabilities of 
the system.
In our results if $\alpha_{Lattice} < 0$ means the negative coefficient of the 
kinetic term (\ref{eq:scalarP}), and the stability of the surfaces certainly 
disappear.

\begin{figure}
\vspace*{-0.5cm}
\centerline{
\psfig{file=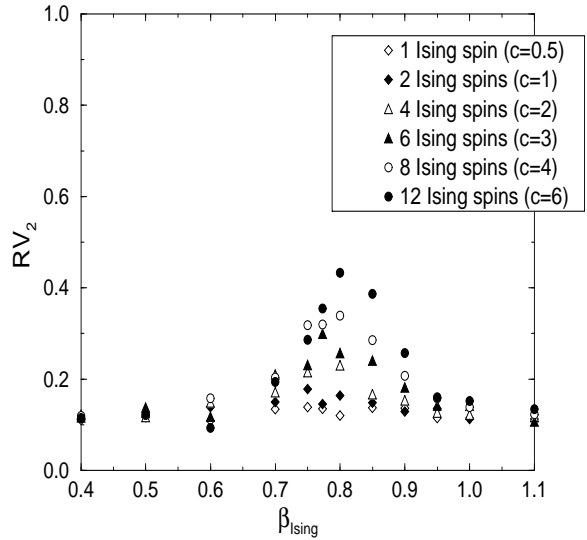,height=8cm,width=8cm}} 
\caption
{
$RV_{2}$ plotted versus the coupling constant of the Ising spins 
($\beta_{Ising}$).
The open diamond indicates one Ising spin($c=0.5$), the close diamond 
indicates two Ising spins($c=1$), the open triangle indicates 4-Ising 
spins($c=2$), the close triangle indicates 6-Ising spins($c=3$), the open 
circle indicates 8-Ising spins($c=4$) and the close circle indicates 
12-Ising spins($c=6$). 
The total number of triangles is 2,000, and averaged by 500 
ensembles for each data.
}
\label{fig:IsingRV}
\vspace{-0.5cm}
\end{figure}

\begin{figure}
\vspace{-0.5cm}
\centerline{
\psfig{file=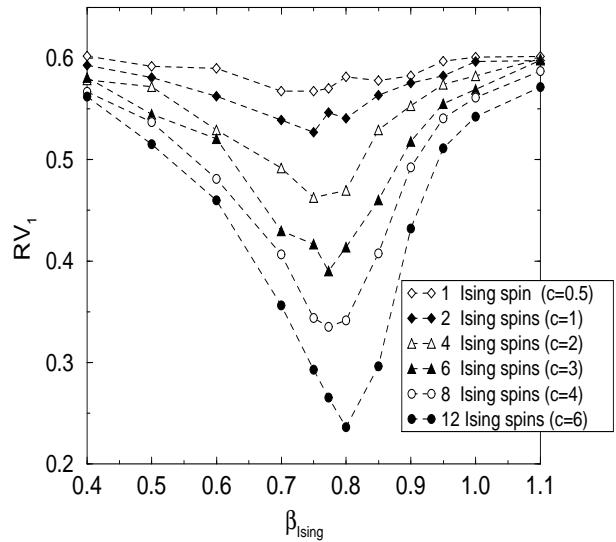,height=8cm,width=8cm}} 
\caption
{
$RV_{1}$ ($V_{total}=2000$) plotted versus $\beta_{Ising}$; the other 
conditions are the same as in Fig.12.
}
\label{fig:IsingMaxVol}
\vspace{-0.5cm}
\end{figure}

\begin{figure}
\vspace{0cm}
\centerline{
\psfig{file=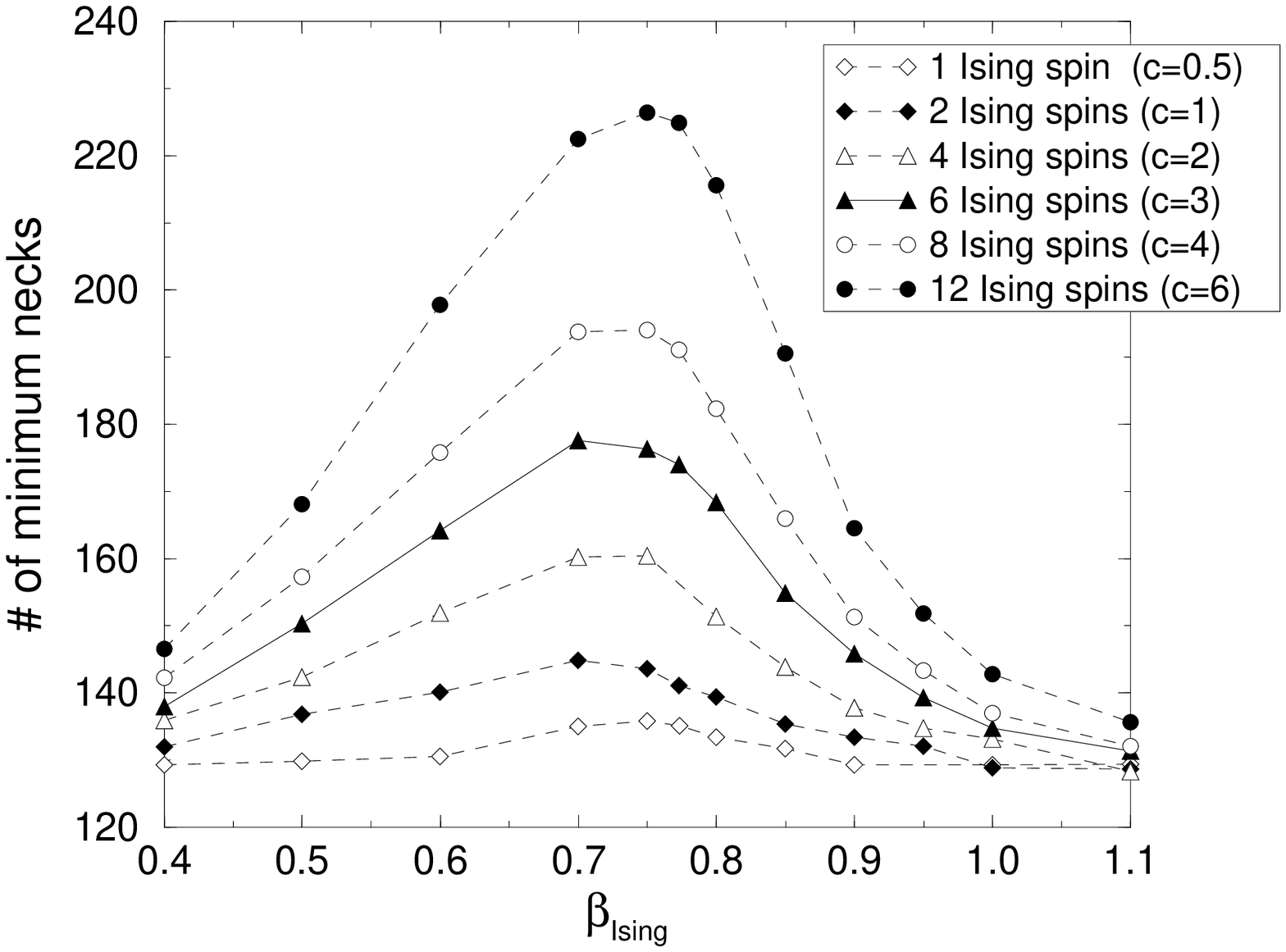,height=8cm,width=8cm}} 
\caption
{
Total number of minimum necks plotted versus $\beta_{Ising}$; the other 
conditions are the same as in Fig.12. 
}
\label{fig:IsingMinbuNecks}
\vspace{0.0cm}
\end{figure}

\subsection{Matter($c>1$) coupled with gravity}
There are many theoretical indications that the surfaces degenerate 
into branched-polymer configurations for $c>1$.
We investigated the case of multi-Ising spins coupled to gravity using 
the minbu analysis.
Each Ising spin lives on a triangles, and is thermalized by using cluster 
updates\cite{Wolff}

It is rather difficult to obtain clear signals for such transitions 
when n-Ising spins are coupled to gravity.
In Figs.\ref{fig:IsingRV}, \ref{fig:IsingMaxVol} and 
\ref{fig:IsingMinbuNecks} we plot $RV_{2,1}$ and the averaged total number 
of minimum necks of the surface as a function of the coupling strength of 
the Ising spins, respectively.
All of these observables slowly change with the central charge; however, 
it is certain that the effect of matter makes the surfaces branched for a 
sufficiently large $c$.

It seems natural that the Ising spins living on a small-size baby universes, 
or branched polymers, suffer considerably from finite size effects.
It is thus expected that such spins decouple with the background geometries.
This is the main reason that the transition is not very clear.

\section{Summary and Discussion}
We have classified the two-dimensional random surfaces generated by dynamical 
triangulation into three states (crumpled surface, flat surface and branched 
polymer) varying the strength of two interaction terms.
The three states are characterized by local observables, such as average 
square curvatures ($Q_{2}$) and the average square-curvature derivative 
($Q_{3}$).
The $R^{2}-$gravity in the strong coupling limit $\frac{\beta}{A} \to \infty$, 
gives the flat-surface, which was established numerically. 
We could also explain the cross-over phenomenon between the fractal and flat 
states within the perturbative colculations.
Furthermore, we examined why the surfaces become crumple rather than a 
branched polymer in the region of $ \alpha_{Lattice} > 0 , 
\beta_{Lattice} < \alpha_{Lattice}$. 
On the other hand we observed that the surface becomes branched polymers with 
$\alpha_{Lattice} < 0 , \beta_{Lattice} < |\alpha_{Lattice}|$, and that the 
measurements of its resistivities clarify that this surface does not have a 
smooth continuum limit.
The minbu analysis showed the nested structure of random surfaces directly; 
it also gave one important clue for distinguishing branched polymers from 
other kinds of surfaces. 
As to whether a mother universe exists or not, we quantitatively propose an 
observable, $RV_{1} \equiv \frac{V_{max}}{V_{total}}$ or 
$RV_{2} \equiv \frac{V_{second}}{V_{max}}$.
In fact, the $RV_{1,2}$ works well in our simulations.
The good fractal structure which is known in the case of pure gravity
($\alpha \sim \beta \sim 0$) disappears in each of three states; on crumpled 
surfaces large branches can not grow enough, and on branched polymers 
a mother universe disappears, while branching into many small baby universes, 
which means a disappearance of the mother loop in the distribution of the loop 
length.
We also applied the same analysis to the case of multi-Ising spins coupled 
to gravity corresponding to $c = 1,2,4,6,8,12$.
The $RV_{1,2}$ smoothly varies, and the number of minimum necks slowly 
grows with the central charges.

The kinetic terms of the auxiliary field of eqs.(\ref{eq:scalarP}) and 
(\ref{eq:scalarN}) have negative signs for $\alpha_{Lattice} < 0$ within 
which we observed the branched polymer configurations in our numerical 
simulation.
Since we do not know a connection between the effective theory coupled to a 
scalar field and the numerical simulation clearly, the central charge 
($c$) which corresponds to the DT surfaces cannot be estimated precisely.
However, we can be fairly certain that these branched polymers are related to 
the $c>1$ instabilities of the surfaces.
We therefore need to inquire further into the intimate correspondences 
between the numerical simulations and continuum theory.

\Large
\begin{center}
Acknowledgements
\end{center}
\normalsize
We would like to thank H.Hagura, H.Kawai, H.S.Egawa, J.Nishimura, N.Ishibashi 
and S.Ichinose for useful discussions and comments. 
One of the authors (N.T.) was supported by a Research Fellowship of the Japan 
Society for the Promotion of Science for Young Scientists.

\Large
\begin{center}
Appendix
\end{center}
\normalsize
\appendix
\section{Semi-classical calculation of $R^{2}$-gravity}
\def\Dg{{\cal D}_{\hat g}}
\def\D{{\cal D}}
\def\mn{{\mu\nu}}
\def\e{{\hbox{e}}}
In this appendix we consider a semiclassical treatment based on a ref.
\cite{Zamo} of $R^{2}$-gravity in order to account for the cross-over 
transition between a flat surface and a fractal surface.

The partition function for surfaces with a fixed area of $A$ is defined by
$$
Z[A] = \int{\D g_{\mn}\D X \over \hbox{\it Vol.(Diffeo)}} 
\e^{-S_G[g]-S_m[X]}\delta\left(\int dx^2\sqrt{g}
-A\right),
\eqno{(A.1)}
$$
where 
$
S_{G}[g] = \int d^2x \sqrt{g}(\lambda + \beta R^{2}),
$
and $S_m[X]$ is the matter action given by
$
S_m[X]=\int d^2x\sqrt{g}g^\mn\partial_\mu X^a\partial_\nu X^a,
\quad a=1 \sim d.
$

Taking the conformal gauge $g_{\mn}=\hat g_{\mn}e^{\phi}$, 
and estimating the FP-determinant and the integral of matter 
fields, we have
$$
Z(A)=\int{\D_g\phi\over\hbox{\it Vol.(C.G.)}} 
\e^{-S_G[\hat g\e^{\phi}]-
{26-d\over96\pi}\int d^2x\sqrt{g}(-\phi\Delta_1\phi+2R\phi)}
\delta\left(\int d^2x\sqrt{\hat g}\e^{\phi}-A\right),
\eqno{(A.3)}
$$
where $\Delta_{1}$ is the Laplacian of $S^{2}$ with radius $1$.
Then, integrating the constant mode of $\phi$, which is only used to 
estimate the $\delta$-function, as
$\phi_0=\ln A - \ln{G[\tilde{\phi}]}$, where 
$G[\tilde{\phi}] = \int d^2x\sqrt{\hat g}\e^{\tilde\phi}$, we obtain
$$
Z(A,\beta)=A^{d-20\over6}\e^{-\lambda A}
\int\Dg\tilde\phi\e^{-\tilde S[\tilde\phi]},
\eqno{(A.4)}
$$
where the action for oscillating parts, $\tilde\phi=\phi-\phi_0$, is now
$$
\tilde S[\tilde\phi]=\int d^2x\sqrt{g_1}\left(
{\beta G[\tilde\phi]\over Ar^2}(2-\Delta_1\tilde\phi)^2\e^{-\tilde\phi}
-{1\over2\gamma}\tilde\phi\Delta_1\tilde\phi
-{2\over\gamma}\ln G[\tilde\phi]\right).
\eqno{(A.5)}
$$
Here we have written $\gamma={48\pi\over 26-d}$.
Now, we make a weak field expansion, and take only the quadratic terms 
with respect to the field, $\tilde\phi$.
The functional integral is carried out by expanding $\phi$ in terms of 
spherical-harmonic functions. 
Finally, we obtain the expression
$$
Z(A,\beta)=A^{d-20\over6}
\e^{-\lambda A-{64\pi^2\beta\over A}}
\prod_{l\ge2}^L(\omega_l)^{-{2l+1\over2}}, 
\eqno{(A.6)}
$$
where
$$
\omega_l={(4\pi)^2\beta\over A^2}(l(l+1)-2)^2+
{2\pi\over\gamma}{(l(l+1)-2)\over A}.
$$
The zero-frequency modes with $l=1$ correspond to the 
conformal Killing modes, and have been omitted in the 
functional integral. 

For a large $\beta$, the determinant part is 
$$
\matrix{
\sum_{l\ge2}^L(2l+1)\ln\omega_l
\sim&\left(2L(L+2)-{8\over3}+{A\over8\pi\beta\gamma}\right)
\ln L^2-2L^2 \hfill\cr
&-2(L(L+2)-3)\ln A+(L(L+2)-3)\ln(16\pi^2\beta).\hfill
}
\eqno{(A.7)}
$$

The ultra-violet cut-off of an angular momentum ($L$) will be fixed by 
$L(L+2) = A \Lambda^2$.
We obtain
$$
\sum_{l\ge2}^L(2l+1)\ln\omega_l\sim\kappa_0+\kappa_1(\beta)A+
\left({10\over3}+{A\over8\pi\beta\gamma}\right)\ln A
-3\ln(16\pi^2\beta),
\eqno{(A.8)}
$$
where $\kappa_0$ and $\kappa_1$ are divergent when 
the cut-off ($\Lambda$) goes to infinity.

For small $\beta$, we have
$$
\sum_{l\ge2}^L(2l+1)\ln\omega_l\sim\kappa_0'+\kappa_1'(\beta)A+
{5\over3}\ln A + 3\ln\left({\gamma\over2\pi}\right).
\eqno{(A.9)}
$$

In order to compare the semi-classical calculation with the numerical 
results, we should make a correspondence of the cut-off ($\Lambda$) with 
$N_{2}$ as $L(L+2) = A \Lambda^{2} = \kappa N_{2}$, where $\kappa$ is a 
constant which should be determined based on the numerical data. 
Using the asymptotic formulas eqs.$(A.8)$ and $(A.9)$, we obtain 
$
\langle \int d^{2}x R^{2} \rangle = - \frac{\partial \log Z}{\partial \beta}
$
for large $\beta_{Lattice}$,
$$
a^{2} \left\langle \int \sqrt{g} R^{2} d^{2}x \right\rangle
= 48\pi^{2}N_{2} 
\left\{ 
       \frac{\kappa}{2\beta_{Lattice}} + \frac{4}{3N^{2}_{2}}
       + \frac{6\pi \ln({\kappa N_{2}})}{\beta_{Lattice}^{2}\gamma}
\right\},
\eqno{(A.10)}
$$
and for small $\beta_{Lattice}$
$$
a^{2} \left\langle \int \sqrt{g} R^{2} d^{2}x \right\rangle
= 12\pi N_{2} \gamma 
\left\{ 
        \frac{\kappa^{2}}{6} + \frac{1}{N_{2}}
\right\}.
\eqno{(A.11)}
$$
We are able to determine the value of $\kappa$ from a data point, for 
example $(Q_{2},\beta_{Lattice})$ = $(0.0449 \pm 0.0005,200)$.
From eqs.(\ref{eq:RR}),(\ref{eq:corresp}),(\ref{eq:Q2_def}) and ($A.10$) 
we obtain 
$$
\kappa = 0.499 \pm 0.005,
\eqno{(A.12)}
$$
which is consistent with ref.\cite{Tsuda_Yukawa}.

From eq.($A.6$) we can calculate $\langle\int d^2x\sqrt{g}$ $R^{2}$ 
$\rangle$ without using the asymptotic formulas given above.
We thus have
$$
{\left\langle \int d\Omega R^2\right\rangle\over48\pi^2 N \Lambda^2}=
{4\over3N^2}+
{1\over N}\sum_{l\ge2}^{ \sqrt{\kappa N}}{2l+1\over2}
{{1\over N}(l(l+1)-2)\over{\beta_{Lattice}\over N}(l(l+1)-2)+{6\pi\over\gamma}},
\eqno{(A.12)}
$$
which can well reproduce the data (Fig.\ref{fig:R2}) in both ranges of 
small and large $\beta$, when we choose the parameter $\kappa$ to be 0.5. 

The values of $\gamma_{string}$ given through semi-classical calculation, 
$$
\gamma_{string}=\left\{\matrix{{d-7\over6}\hfill&\quad
\hbox{ for small }\beta\cr
{d-12\over6}+{(d-26)A\over768\pi^2\beta}\hfill&
\quad\hbox{ for large }\beta,
}\right.
\eqno{(A.13)}
$$
is exact in a large (-$d$) limit. 
If one wants to obtain correct values for all $d$, a full quantum treatment 
is necessary.


\end{document}